\documentclass[12pt]{article}

\usepackage[T1]{fontenc}
\usepackage[utf8]{inputenc}
\usepackage[margin=1in]{geometry}
\usepackage{graphicx}
\usepackage{xcolor}
\usepackage{bm}
\usepackage{mathtools}
\usepackage{amssymb}
\usepackage{placeins}
\usepackage[authoryear,round]{natbib}
\usepackage[hidelinks]{hyperref}
\graphicspath{{figs/}}

\newcommand{\eq}[1]{\begin{align}#1\end{align}}
\newcommand{\pd}{\partial}
\newcommand{\nb}{\nabla}
\newcommand{\nn}{\nonumber}

\numberwithin{equation}{section}
\makeatletter
\@ifundefined{varGamma}{\DeclareMathSymbol{\varGamma}{\mathord}{letters}{"00}}{}
\@ifundefined{varLambda}{\DeclareMathSymbol{\varLambda}{\mathord}{letters}{"03}}{}
\@ifundefined{varPi}{\DeclareMathSymbol{\varPi}{\mathord}{letters}{"05}}{}
\@ifundefined{varSigma}{\DeclareMathSymbol{\varSigma}{\mathord}{letters}{"06}}{}
\@ifundefined{varOmega}{\DeclareMathSymbol{\varOmega}{\mathord}{letters}{"0A}}{}
\makeatother

\title{The critical radius of compressible capillary drops: 
viscosity, thermodynamics, and diffuse-interface scales}

\author{Umpei Miyamoto\\[5pt]
\small Research and Education Center for Comprehensive Science\\[1pt]
\small Akita Prefectural University, Akita 015-0015, Japan\\[3pt]
\small \href{mailto:umpei@akita-pu.ac.jp}{umpei@akita-pu.ac.jp}}

\date{}

\begin{document}

\maketitle

\begin{abstract}
A compressible capillary drop differs from an incompressible one in that its radius is a dynamical degree of freedom.  When the drop is sufficiently small, this spherical mode can become unstable, defining a critical radius set by the competition between surface tension and compressibility.  This paper examines the robustness and physical meaning of that critical radius.  For a non-relativistic viscous fluid, starting from the viscous compressible equations and the free-surface stress condition, we derive the radial dispersion relation and show that shear and bulk viscosities change the eigenvalues but not the onset radius.  A thermodynamic argument identifies the same radius as the point where the energy of a uniformly compressed drop changes from locally stable to unstable, explaining why the threshold is not set by viscous dissipation.  This energetic interpretation can also be applied to a special-relativistic fluid, where the non-relativistic mass-density factor is replaced by the corresponding enthalpy-density factor.  We then compare the critical radius with diffuse-interface length scales in two Cahn--Hilliard free-energy models to determine whether the instability can occur within the range of validity of the sharp-interface description.  In a symmetric quartic model the critical radius is much smaller than the interface thickness, so the instability is absent throughout that range.  In a shallow-well model, however, the critical radius can become parametrically larger than the interface thickness, leaving a range of sharp-interface drops that are unstable.  Whether such a range exists therefore depends on the diffuse-interface free-energy model.
\end{abstract}

\section{Introduction}
\label{sec:introduction}

The classical theory of capillary drops is built around the motion of a free surface.  For an incompressible drop, the normal modes studied by \citet{Kelvin}, \citet{RayleighSound} and \citet{Lamb} are shape oscillations at fixed volume.  Viscosity, damping and related corrections were later analysed by \citet{Chandrasekhar}, \citet{Reid} and \citet{Becker}.  The same capillary balance also underlies the Plateau--Rayleigh instability of liquid cylinders \citep{Plateau,Rayleigh,ChandrasekharBook,Eggers:1997zz}.  These classical problems establish the standard capillary spectrum and instability mechanisms, but the incompressibility assumption removes the uniform radial, or breathing, degree of freedom of a drop.

The spherical mode is absent for an incompressible liquid, because a uniform radial motion would change the volume at fixed mass.  For a compressible drop, however, the radius is a genuine dynamical degree of freedom.  This mode changes the volume while preserving the spherical shape, so the pressure response and the capillary pressure determine whether the equilibrium radius is restoring or destabilizing.  \citet{Miyamoto:2011hr} studied the perfect-fluid version of this problem and found that a sufficiently small compressible drop is unstable.

{\color{black}To the best of our knowledge, \citet{Miyamoto:2011hr} provided the
first explicit normal-mode analysis that identified a
capillary--compressibility threshold for the spherical mode of a classical
compressible drop.}

{\color{black}Earlier thermodynamic work had identified a capillary-compression
length given by the product of isothermal compressibility and surface tension,
together with the associated static shrinkage of a spherical drop
\citep{EgelstaffWidom1970,Dixmier2006}.  Those results did not interpret this
length as a stability threshold or identify an unstable radial mode.}

{\color{black}Compressible viscous capillary fluids with free boundaries have also
been studied in the mathematical literature; for example,
\citet{Zajaczkowski1994} proved global existence near a spherical domain.
Such existence results do not isolate the spherical
capillary--compressibility mode or derive its critical radius.}

{\color{black}Related thresholds have since appeared in distinct settings.
\citet{TamimBostwick2019} found a compressibility-driven volume instability
of an elastic gel drop, \citet{Villemot2020} obtained a static collapse
criterion for a two-dimensional compressible droplet, and
\citet{Mitsuhashi2026} recently identified a mechanically unstable breathing
mode of a self-bound superfluid droplet.}

{\color{black}These models differ from the classical Newtonian liquid considered
here.  To the best of our knowledge, the effects of Newtonian shear and bulk
viscosities on the present instability have not previously been analysed.}

For the non-relativistic perfect fluid, let \(\rho_0\) be the equilibrium mass density and \(c_s\) the adiabatic sound speed at equilibrium.  For a drop bounded by an \(n\)-sphere with surface tension \(\sigma\), the critical radius is
\eq{
	{\color{black}R_c}
	=
	\frac{n}{n+1}\frac{\sigma}{\rho_0 c_s^2}.
}
A non-relativistic drop with radius below \({\color{black}R_c}\) is unstable to the spherical perturbation.

For comparison, the special-relativistic perfect-fluid result can be stated separately.  In units \(c=1\), let \(\mathcal E_0\) be the equilibrium energy density and let \(c_s\) now denote the relativistic sound speed.  The critical radius is
\eq{
	{\color{black}R_{c,{\rm SR}}}
	=
	\frac{n}{n+1}\frac{\sigma}{\mathcal E_0 c_s^2}
	\left[1-(n+1)c_s^2\right].
}
This expression applies when \(c_s^2<1/(n+1)\).  There is no positive critical radius of this type when \(c_s^2\geq1/(n+1)\).  {\color{black}The subscript distinguishes this special-relativistic radius from the non-relativistic \(R_c\).}

In this paper, we revisit this spherical compressible-drop instability for a non-relativistic viscous fluid.  Viscosity should certainly change the time-dependent roots of the dispersion relation, because it damps motion and also enters the normal stress at the surface.  The question is whether it also changes the critical radius.  We derive the linearized viscous compressible problem for a spherical perturbation and obtain the characteristic equation.  The answer is simple: shear and bulk viscosities do not move the critical radius.  They affect the local behaviour of the root near the marginal point, but the location of that point is unchanged.

We then give a thermodynamic interpretation of this result.  At fixed mass and entropy, the threshold is the point where the second derivative of the internal-plus-surface energy with respect to the radius changes sign.  Viscosity is not part of this static energy, so it cannot change this radius.  The same energy argument also rederives the special-relativistic critical radius.  The polytropic case is included as a simple example in which the full radius-dependent energy can be plotted explicitly.

{Whether the instability can occur within the continuum regime is
a separate question.  For water in air at \(25\,{}^\circ\mathrm{C}\), the representative values \(n=2\),
\(\sigma=72.0\times10^{-3}\,\mathrm{J\,m^{-2}}\),
\(\rho_0=1.00\times10^3\,\mathrm{kg\,m^{-3}}\) and
\(c_s=1.50\times10^3\,\mathrm{m\,s^{-1}}\) give
\(R_c\simeq2.13\times10^{-11}\,\mathrm{m}\).  As already emphasized by
\citet{Miyamoto:2011hr}, this value is far below the molecular length scales
required for a continuum description.}

{The water estimate therefore motivates a diffuse-interface
comparison of the critical radius with the interface thickness, since the
critical radius must be much larger than the interface thickness for a range of
unstable drops to lie within the sharp-interface regime.  In the final part of
the main text, we compare the critical radius and the interface thickness in
the Cahn--Hilliard framework by deriving surface tension, compressibility and
interface thickness from a single free-energy functional and applying the
resulting static relations to two free-energy models
\citep{CahnHilliard,Rowlinson}.}

For a symmetric quartic potential, every drop large enough for the
sharp-interface approximation lies above the critical
radius and is therefore stable.  A shallow-well model, however, leaves a
range of unstable drops whose radii are still much larger than the interface
thickness.

The appendices clarify the scope of the result.  Appendix~\ref{app:mass-conservation} records how the continuity equation and the kinematic free-boundary condition imply conservation of total mass in covariant spatial notation.  Appendix~\ref{app:nonspherical-spectrum} recalls the perfect-fluid nonspherical spectrum of \citet[Appendix B, equation (B13)]{Miyamoto:2011hr}, illustrating that nonspherical capillary frequencies decrease with compressibility but do not turn into unstable modes in that problem.  Appendix~\ref{app:sr-thermo} repeats the energy argument for a special-relativistic perfect fluid and shows that the non-relativistic inertial factor is replaced by the enthalpy density.  Appendix~\ref{app:surface-gravity-gamma} contrasts the surface-tension result with the familiar homologous estimate for Newtonian self-gravity.

The paper is organized as follows.  Section~\ref{sec:setup} sets up the viscous compressible equations and boundary conditions.  Section~\ref{sec:linear} derives the radial perturbation equation and characteristic equation, then nondimensionalizes the result.  Section~\ref{sec:viscosity} locates the critical compressibility from the small-growth-rate expansion, illustrates the root structure and discusses the near-threshold root scalings.  Section~\ref{sec:thermo} gives the energy argument, including the polytropic example.  Section~\ref{sec:diffuse} reviews the diffuse-interface length scales needed for the comparison and applies them to two Cahn--Hilliard models, and Section~\ref{sec:conclusion} summarizes the main points.

\section{Viscous compressible fluid with free boundary}
\label{sec:setup}

We review the Navier--Stokes equation and its boundary conditions \citep[see e.g.][]{Landau:1987gn}, while generalizing them to arbitrary dimensions and curvilinear coordinates. 

\subsection{Equations of motion and boundary conditions in general form}
\label{sec:NS}

We consider a fluid in a $d$-dimensional flat spacetime $\mathbb{R}^{1,d-1}$ ($d \geq 4$) with time coordinate $t$ and general curvilinear spatial coordinates $x^I$ ($I,J=1,2,\ldots,d-1$).  The spatial metric is time independent, and the equations below are written in a form covariant under changes of the spatial coordinates.  The continuity equation is
\eq{
	\pd_t \rho + \nb_I ( \rho v^I ) = 0,
\label{eq:eoc-1}
}
where $\rho$ is the density, $v^I$ the fluid velocity, and $\nb_I$ the covariant derivative compatible with the spatial metric \(g_{IJ}(x)\) \citep[see e.g.][]{Wald}.
The equations of motion follow from momentum conservation,
\eq{
	\pd_t ( \rho v^I ) + \nb_J ( \rho v^I v^J - \varPi^{IJ} ) = 0,
\label{eq:mom-cons}
}
where $\varPi_{IJ}$ is the stress tensor.  For a non-relativistic viscous fluid we take
\eq{
	\varPi_{IJ}
	=
	- p g_{IJ}
	+
	\eta \left(
		\nb_{I} v_{J} + \nb_{J} v_{I}
		-
		\frac{2}{d-1} g_{IJ} \theta
	\right)
	+
	\zeta g_{IJ} \theta,
\qquad
	\theta \coloneqq \nb_I v^I,
\label{eq:Pi}
}
where $p$ is the pressure, $\eta$ and $\zeta$ are the shear and bulk viscosities, respectively, and $\theta$ is the expansion.  The viscosity coefficients are assumed to be non-negative constants.
Substituting equation~\eqref{eq:Pi} into equation~\eqref{eq:mom-cons} and using equation~\eqref{eq:eoc-1} gives the Navier--Stokes equation
\eq{
	\rho
	( \pd_t + v^J \nb_J ) v_I
	=
		- \nb_I p + \eta \Delta v_I 
	+ \left(  \zeta + \frac{ d-3 }{ d-1 } \eta \right)
	\nb_I \theta,
\quad
		\Delta \coloneqq \nb^J \nb_J.
\label{eq:NS-1}
}

We next impose free-boundary conditions for two immiscible fluids, denoted by fluids 1 and 2, separated by an infinitesimally thin surface.  The surface is specified by the vanishing of a scalar function, $\varphi(t,x^I)=0$.  Denoting the unit normal by
\eq{
	n_I = \frac{\nb_I \varphi}{|\nabla\varphi|},
	\qquad
	|\nabla\varphi|
	\coloneqq
	\sqrt{\nb^J \varphi\,\nb_J \varphi},
}
pointing from fluid 1 to fluid 2, and denoting the surface tension by $\sigma(t,x^I)$, the stress balance condition is the Young--Laplace relation
\eq{
	\left(
		\varPi^{(2)}_{IJ} - \varPi^{(1)}_{IJ}
	\right) n^J
	=
	\left.
	\sigma \kappa n_I + h_{I}^{J} \nb_J \sigma
	\right|_{\varphi=0},
\label{eq:YL-1}
}
where $\kappa$ is \((d-2)\) times the mean curvature of the surface and $h_{IJ}$ is the projection tensor,
\eq{
	\kappa
	\coloneqq \nb_I n^I,
\;\;\;
	h_{IJ} \coloneqq g_{IJ} - n_I n_J.
\label{eq:kappa-1}
} 

The kinematic free-surface condition states that the surface moves with the fluid,
\eq{
	 ( \pd_t + v^J \nb_J ) \varphi = 0  \big|_{\varphi=0}.
\label{eq:kinetic-1}
}
Together with equation~\eqref{eq:eoc-1}, this material-boundary condition implies conservation of the total mass of the drop; see Appendix~\ref{app:mass-conservation}.
This check is used later in section~\ref{sec:thermo}, where the energetic
interpretation is formulated as a fixed-mass variation of the equilibrium
radius.

We assume that the surface tension $\sigma$ is constant and neglect the stress of fluid 2, $\varPi^{(2)}_{IJ}=0$.  Omitting the label of fluid 1, the normal and tangential components of equation~\eqref{eq:YL-1} are then
\eq{
	-  n_I \varPi^{IJ}  n_J
	&=
	\sigma \kappa  \big|_{\varphi=0},
\label{eq:YL-norm-1}
\\
	h_{I}^{J} \varPi_{JK} n^K &= 0  \big|_{\varphi=0}.
\label{eq:YL-tan-1}
}
\subsection{Equations under spherical symmetry}
\label{sec:axial}

For spherically symmetric flows it is convenient to introduce a radial coordinate system in which the line element of \((d-1)\)-dimensional flat space is
 \eq{
	g_{IJ} d x^I d x^J
	=
	d r^2 + r^2 d s_n^2
	=
	dr^2 + r^2 \varOmega_{ij}(\vartheta) d \vartheta^i d \vartheta^j,
\qquad
	n \coloneqq d-2,
\label{eq:metric}
}
where $ d s_n^2 = \varOmega_{ij}(\vartheta) d \vartheta^i d \vartheta^j \; (i,j=1,2,\dots,n)$ is the line element of a unit $n$-sphere.
The independent nonzero Christoffel symbols of equation~\eqref{eq:metric} are
\eq{
	\varGamma^r_{ij}=-r\varOmega_{ij},
	\qquad
	\varGamma^i_{rj}=\frac{1}{r}\delta^i_j,
	\qquad
	\varGamma^i_{jk}={}^{(\varOmega)}\varGamma^i_{jk},
\label{eq:christoffel-sphere}
}
where ${}^{(\varOmega)}\varGamma^i_{jk}$ is the Christoffel symbol of the unit $n$-sphere metric $\varOmega_{ij}$.

We assume that the velocity is purely radial and that the free surface is spherically symmetric:
\eq{
	v^r = v(t,r),
\;\;\;
	v^i = 0 \;\; (i=1,2,\dots,n),
\;\;\
	\varphi = r - R(t),
\label{eq:axial}
}
where $R(t) \; (>0)$ represents the instantaneous radius of a drop.

With this ansatz, the continuity equation~\eqref{eq:eoc-1} and the \(r\)-component of equation~\eqref{eq:NS-1} become
\eq{
	&\left(
		\pd_t + v \pd_r
	\right) \rho
	+
	\rho \theta
	=0,
\qquad
	\theta
	=
	\left( \pd_r + \frac{n}{r} \right) v.
\label{eq:eoc-2}
\\
	\rho
	\left(
		\pd_t + v \pd_r
	\right) v
	&=
	- \pd_r p
	+ \eta
	\left(
		\pd_r^2 + \frac{n}{r} \pd_r 
		- \frac{n}{r^2} 
	\right) v
	+
	\left(
		\zeta + \frac{n-1}{n+1} \eta
	\right)
	\pd_r \theta.
\label{eq:NS-2}
}
The normal component of the stress-balance equation~\eqref{eq:YL-norm-1} is given by
\eq{
	p
	-
	2\eta
	\pd_r v
	- \left( \zeta - \frac{2}{n+1} \eta \right) \theta
	=
	\sigma \kappa  \big|_{r=R},
\qquad
	\kappa
	=
	\frac{n}{r}.
\label{eq:YL-norm-2}
}
Under spherical symmetry the tangential stress balance~\eqref{eq:YL-tan-1} is identically satisfied.  The kinematic boundary condition~\eqref{eq:kinetic-1} reduces to
\eq{
	\pd_t R = v \big|_{r=R}.
\label{eq:kinetic-2}
}

\section{Linear perturbation of a static drop}
\label{sec:linear}

\subsection{Perturbed equations}

The above system admits a static drop solution parametrized by
\eq{
	R={\color{black}R_0},
\;\;\;
	p=p_0 \coloneqq \sigma \frac{n}{{\color{black}R_0}},
\;\;\;
	\rho = \rho_0 \coloneqq \rho (p_0),
\;\;\;
	v = 0,
\label{eq:bg}
}
where ${\color{black}R_0}$ is the radius of the drop and we have assumed a barotropic equation
of state, evaluated locally as \(\rho=\rho(p)\) near the equilibrium pressure
\(p_0\).

We perturb the static solution as
\eq{
	R={\color{black}R_0} + \delta R(t),
\;\;\;
	p = p_0 + \delta p(t,r),
\;\;\;
	\rho = \rho_0 + c^{-2}_s \delta p(t,r),
\;\;\;
	v = \delta v(t,r),
\label{eq:linear}
}
where
\eq{
	c_s^2 \coloneqq \left. \frac{ dp }{ d\rho } \right|_{\rho=\rho_0},
\label{eq:cs2}
}
is the squared sound speed at the equilibrium state.

Substituting equation~\eqref{eq:linear} into the continuity equation~\eqref{eq:eoc-2} and the Navier--Stokes equation~\eqref{eq:NS-2},
and retaining the terms linear in the perturbations, we obtain
\eq{
	&
	\pd_t \delta p + c_s^2 \rho_0 \delta \theta = 0,
\qquad
	\delta \theta
	=
	\left( \pd_r + \frac{n}{r} \right) \delta v,
\label{eq:eoc-3}
\\
	\rho_0 \pd_t \delta v
	&=
	-\pd_r \delta p
	+ \eta \left(
		\pd_r^2  + \frac{n}{r} \pd_r
		- \frac{n}{r^2}
	\right) \delta v + \left( \zeta + \frac{n-1}{n+1}\eta \right) \pd_r \delta\theta.
\label{eq:NS-3}
}
Linearizing the Young--Laplace relation~\eqref{eq:YL-norm-2} and the kinematic boundary condition~\eqref{eq:kinetic-2} gives
\eq{
	\delta p - 2\eta \pd_r \delta v
	-\left( \zeta-\frac{2}{n+1} \eta \right) \delta \theta
	&=
	\sigma \delta \kappa  \big|_{r={\color{black}R_0}},
\label{eq:YL-norm-3}
\qquad
	\delta \kappa
	=
	-\frac{n}{{\color{black}R_0}^2} \delta R,
\\
	\pd_t \delta R
	&=
	\delta v  \big|_{r={\color{black}R_0}}.
\label{eq:kinetic-3}
}

\subsection{Separation of variables}

We separate variables by writing the perturbations as
\eq{
	\delta R = {\color{black}R_0} e^{\omega t},
\;\;\;
	\delta p = P(r) e^{\omega t},
\;\;\;
	\delta v= u (r) e^{\omega t},
\label{eq:sov}
}
where \(\omega\) is an unknown constant, and \(P(r)\) and \(u(r)\) are unknown functions to be determined.  {\color{black}An arbitrary infinitesimal constant multiplying all three perturbations has been suppressed.}  The common factor \(e^{\omega t}\) is divided out below.

Define two viscosity combinations
\eq{
	\zeta_+ \coloneqq \zeta+\frac{2n}{n+1}\eta,
	\qquad
	\zeta_- \coloneqq \zeta-\frac{2}{n+1}\eta.
}
The perturbed continuity equation~\eqref{eq:eoc-3} and Navier--Stokes equation~\eqref{eq:NS-3} then give
\eq{
	&
	\omega P
	+ c_s^2 \rho_0 
	\left( \frac{d}{dr} + \frac{n}{r} \right) u = 0,
\label{eq:eoc-4}
\\
	&
	\omega \rho_0 u
	=
	- \frac{dP}{dr} 
	+ \zeta_+ \frac{d}{dr} \left( \frac{d}{dr} + \frac{n}{r} \right) u.
\label{eq:NS-4}
}
The Young--Laplace relation~\eqref{eq:YL-norm-3} and kinematic boundary condition~\eqref{eq:kinetic-3} give
\eq{
	P
	-\left(
		\zeta_+ \frac{d}{dr}  + \zeta_- \frac{n}{r}
	\right) u &= -\sigma \frac{n}{{\color{black}R_0}},
\label{eq:YL-4}
\\
	\omega {\color{black}R_0} & = u,
\label{eq:kinetic-4}
}
where all quantities are evaluated at \(r={\color{black}R_0}\).

\subsection{Solving the perturbation equations}

Eliminating the derivatives of $u$ using equations~\eqref{eq:eoc-4} and \eqref{eq:NS-4}, we obtain
\eq{
	u
	=
	-\frac{\alpha}{\omega\rho_0}\frac{dP}{dr},
\qquad
	\alpha \coloneqq 1+\frac{\zeta_+\omega}{c_s^2\rho_0}.
\label{eq:uP'}
}
Substituting this back into equation~\eqref{eq:eoc-4} gives a second-order ODE for the pressure perturbation:
\eq{
	\frac{d^2P}{dr^2}+\frac{n}{r}\frac{dP}{dr}-\mathcal K^2P=0,
\qquad
	\mathcal K^2 \coloneqq \frac{\omega^2}{\alpha c_s^2}
	=
	\frac{\omega^2}{c_s^2 (1+ \zeta_+ \omega/c_s^2 \rho_0 )},
\label{eq:master-eq}
}
where \(\mathcal K\) plays the role of a radial wave number.  The solution regular at the center \(r=0\) is
\eq{
	P(r)=C\,\frac{\mathrm I_{(n-1)/2}(\mathcal K r)}{(\mathcal K r)^{(n-1)/2}},
\label{eq:P(r)}
}
where \(C\) is a constant and \(\mathrm I_\nu\) denotes the modified Bessel function.
Using the identity
\eq{
		\frac{d}{dx}\left( \frac{\mathrm I_\nu(x)}{x^\nu} \right)
	=
	\frac{\mathrm I_{\nu+1}(x)}{x^\nu},
\label{eq:bessel-identity-derivative}
}
with \(\nu=(n-1)/2\) and \(x=\mathcal K r\), equation~\eqref{eq:u(r)} follows from equations~\eqref{eq:uP'} and \eqref{eq:P(r)}.
\eq{
	u(r)
	=
	-\frac{\alpha C\mathcal K }{\omega\rho_0}\frac{\mathrm I_{(n+1)/2}(\mathcal K r)}{(\mathcal K r)^{(n-1)/2}}.
\label{eq:u(r)}
}
Substituting this into the kinematic boundary condition~\eqref{eq:kinetic-4}, one can fix the constant $C$ as
\eq{
	C
	=
	-\rho_0 c_s^2
	\frac{(\mathcal K {\color{black}R_0})^{(n+1)/2}}{\mathrm I_{(n+1)/2}(\mathcal K {\color{black}R_0})}.
\label{eq:C}
}

Finally, substituting equations~\eqref{eq:P(r)}--\eqref{eq:C} into the Young--Laplace relation~\eqref{eq:YL-4}, and using the derivative identity~\eqref{eq:bessel-identity-derivative} and the recurrence relation
\eq{
	\mathrm I_{\nu+2}(x)
	=
	\mathrm I_{\nu}(x)
	-
	\frac{2(\nu+1)}{x}\mathrm I_{\nu+1}(x),
\label{eq:bessel-identity}
}
one obtains a characteristic equation,
\eq{
	\rho_0\omega^2 {\color{black}R_0}\,
	\frac{\mathrm I_{(n-1)/2}(\mathcal K {\color{black}R_0})}{\mathcal K\,\mathrm I_{(n+1)/2} (\mathcal K {\color{black}R_0})}
	-
	2n\eta\,\omega
	=
	\frac{\sigma n}{{\color{black}R_0}},
\label{eq:disp-1}
}
which encodes the dependence of the perturbation growth rate \(\omega\) on
the seven parameters \(n\), \(\sigma\), \(c_s\), \(\rho_0\), \({\color{black}R_0}\),
\(\eta\), and \(\zeta\).  While the imaginary part of \(\omega\) represents
the oscillatory nature of a stable perturbation, a positive real part of
\(\omega\), if any, implies the existence of an instability of the background
solution~\eqref{eq:bg}.

\subsection{Dimensionless characteristic equation}

To expose the parameters controlling the stability, we introduce the dimensionless quantities
\eq{
	\hat{\omega}
	\coloneqq
	\left(\frac{\rho_0 {\color{black}R_0}^3}{\sigma}\right)^{1/2}\omega,
	\qquad
	\beta
	\coloneqq
	\left(\frac{\sigma}{\rho_0 {\color{black}R_0}}\right)^{1/2}\frac{1}{c_s},
\label{eq:hatomega-beta}
}
together with
\eq{
	\hat{\eta}
	\coloneqq
	\frac{\eta}{(\rho_0 \sigma {\color{black}R_0})^{1/2}},
	\qquad
	\hat{\zeta}
	\coloneqq
	\frac{\zeta}{(\rho_0 \sigma {\color{black}R_0})^{1/2}},
	\qquad
	\hat{\zeta}_{\pm}
	\coloneqq
	\frac{\zeta_{\pm}}{(\rho_0 \sigma {\color{black}R_0})^{1/2}}.
\label{eq:hateta-hatzeta}
}
The parameter \(\beta\) is a dimensionless compressibility parameter.  The parameter \(\hat{\eta}\) is the Ohnesorge number associated with the drop radius; it measures viscous effects relative to inertial and capillary effects.  In the figures below we also use
\(\lambda\coloneqq\zeta/\eta\) when \(\eta>0\) to denote the ratio of bulk to shear viscosity.

With these definitions, equation~\eqref{eq:disp-1} can be rewritten.  From equation~\eqref{eq:master-eq} we have
\eq{
	\mathcal K {\color{black}R_0}
	=
	\frac{\beta \hat{\omega}}{\sqrt{1+\beta^2 \hat{\zeta}_+ \hat{\omega}}}.
\label{eq:calKr0def}
}
Using the definitions above, the characteristic equation~\eqref{eq:disp-1} becomes
\eq{
	\frac{1}{n}\hat{\omega}^2
	\frac{\mathrm I_{(n-1)/2}(\mathcal K {\color{black}R_0})}{\mathcal K {\color{black}R_0}\,\mathrm I_{(n+1)/2}(\mathcal K {\color{black}R_0})}
	-
	2\hat{\eta}\hat{\omega}
	=
	1.
\label{eq:disp-dimless}
}
{\color{black}For nonzero \(\hat{\omega}\), an equivalent characteristic equation is}
\eq{
	G(\hat{\omega})
	\coloneqq
	\hat{\omega}^2
	-
	n\left(1+2\hat{\eta}\hat{\omega}\right)
	\mathcal K {\color{black}R_0}\,
	\frac{\mathrm I_{(n+1)/2}(\mathcal K {\color{black}R_0})}{\mathrm I_{(n-1)/2}(\mathcal K {\color{black}R_0})}
	=
	0,
\label{eq:Ghat}
}
where \(\mathcal K {\color{black}R_0}\) is understood as a function of \(\hat{\omega}\) through equation~\eqref{eq:calKr0def}.  In the inviscid limit \(\hat{\eta}=\hat{\zeta}_+=0\), this reduces to the \(k=0\) perfect-fluid drop equation of \citet[equation (3.1)]{Miyamoto:2011hr}, where \(k\) is the axial wavenumber in the corresponding cylindrical problem.

{\color{black}Equation~\eqref{eq:Ghat} is obtained from
equation~\eqref{eq:disp-dimless} by multiplication by a factor that vanishes
as \(\hat{\omega}^2\) at the origin.  Consequently,
\(G(0)=0\) is a formal zero for every \(\beta\), not by itself a physical
static eigenmode.  We use equation~\eqref{eq:Ghat} for its nonzero roots and
locate marginality by following a nonzero root to the origin.}

We refer to the resulting characteristic equation for \(\omega\) as the dispersion relation, following standard terminology in hydrodynamic stability.

\section{Effect of viscosities on the instability}
\label{sec:viscosity}

\subsection{Critical compressibility}
\label{sec:critical-compressibility}

At a marginally stable static mode, the perturbation neither grows nor decays, so its eigenvalue is \(\hat{\omega}=0\).  A continuous change of stability must therefore be detected by a nontrivial real root approaching the origin.  {\color{black}Accordingly, we examine the nonzero-root limit of equation~\eqref{eq:Ghat} as \(\hat{\omega}\to0\).}  Expanding the Bessel functions and using equation~\eqref{eq:calKr0def}, the characteristic equation becomes
\eq{
\begin{aligned}
	G(\hat{\omega})
	={}&
	\left(1-\frac{n}{n+1}\beta^2\right)\hat{\omega}^2
	+
	\frac{n}{n+1}\beta^2
	\left(\beta^2 \hat{\zeta}_+ - 2\hat{\eta}\right)\hat{\omega}^3
	\\
	&
	+
		\left(
		-\frac{n}{n+1}\beta^6 \hat{\zeta}_+^2
		+\frac{2n}{n+1}\beta^4 \hat{\eta}\hat{\zeta}_+
		+\frac{n}{(n+1)^2(n+3)}\beta^4
		\right)\hat{\omega}^4
		+
		O(\hat{\omega}^5).
	\end{aligned}
	\label{eq:Ghat-smallw}
}
{\color{black}This expansion should first be read as a sign statement near the
origin.  If the coefficient \(1-n\beta^2/(n+1)\) of
\(\hat{\omega}^2\) is negative, then \(G(\hat{\omega})<0\) for
sufficiently small positive real \(\hat{\omega}\).  On the other hand, the
full expression~\eqref{eq:Ghat} gives
\(G(\hat{\omega})\to+\infty\) as \(\hat{\omega}\to+\infty\) along the
positive real axis.  By continuity, \(G\) must then have at least one positive
real root.  Since the perturbation is proportional to \(e^{\omega t}\), this
root is an exponentially growing mode.}

{\color{black}The continuous onset detected by this marginal root is therefore
obtained when the coefficient \(1-n\beta^2/(n+1)\) changes sign.}  This gives
\eq{
	\beta=\beta_c,
	\qquad
	\beta_c
	\coloneqq
	\sqrt{\frac{n+1}{n}},
\label{eq:beta-c1-visc}
}
the same value found in the perfect-fluid calculation of \citet[equation (3.3)]{Miyamoto:2011hr}.  The instability occurs for \(\beta>\beta_c\), namely when the dimensionless compressibility exceeds its critical value.  In terms of the equilibrium radius, equation~\eqref{eq:hatomega-beta} translates this condition into
\eq{
	{\color{black}R_0} < {\color{black}R_c},
	\qquad
	{\color{black}R_c}
	=
	\frac{n}{n+1}\frac{\sigma}{\rho_0 c_s^2}.
\label{eq:rcrit-visc}
}
Viscosities affect the detailed growth rate of this root, but not the critical value at which it appears.

\begin{figure}[!t]
	\centering
	\includegraphics[width=0.72\linewidth]{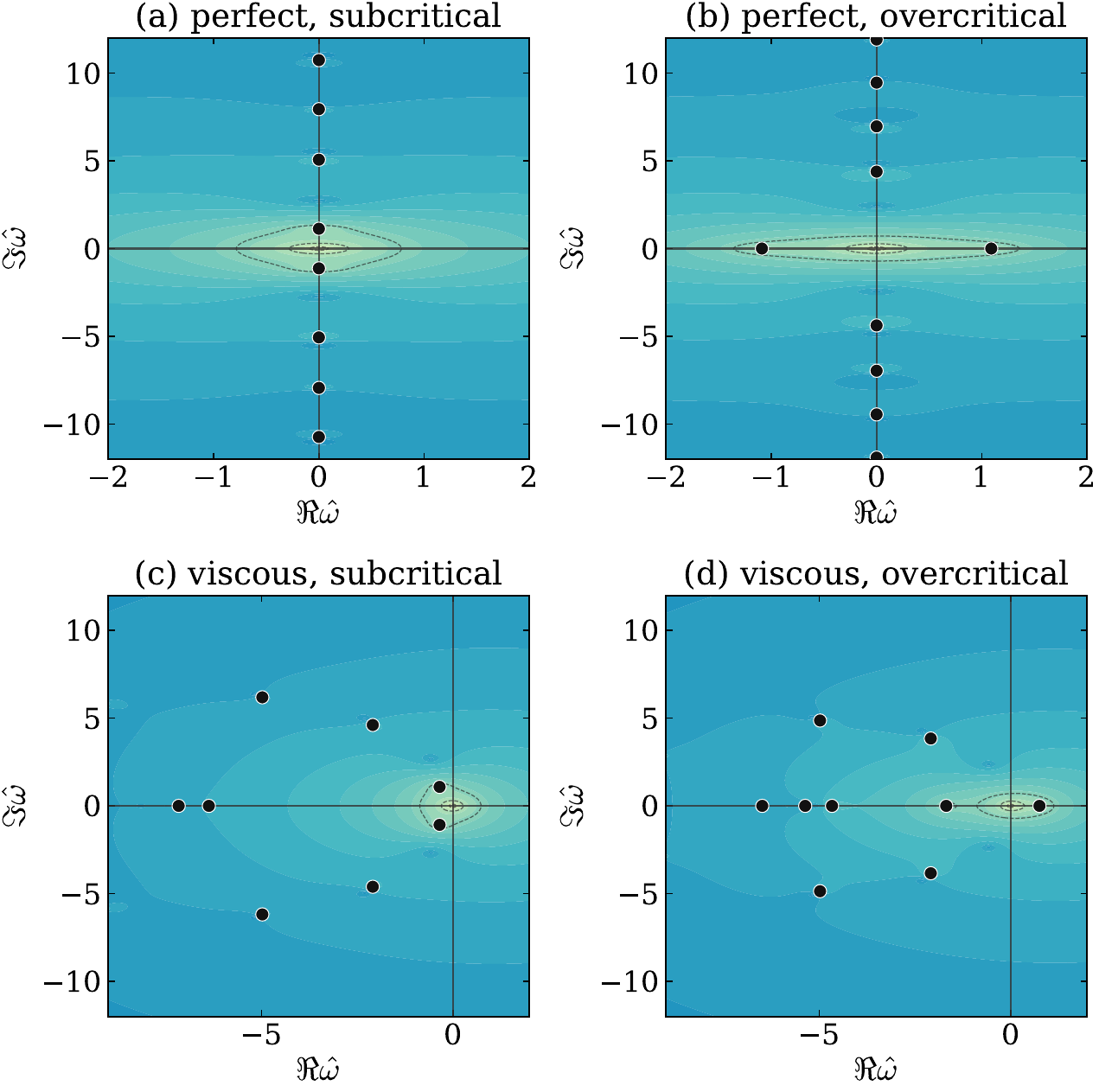}
	\caption{Nonzero roots of \(G(\hat{\omega})=0\) in the complex \(\hat{\omega}\)-plane for \(n=2\).  Black points are zeros, and the background shows contours of \(\log |G|\).  Panels (a) and (c) are subcritical with \(\beta/\beta_c=0.94\), whereas panels (b) and (d) are overcritical with \(\beta/\beta_c=1.06\).  Panels (c) and (d) use \(\hat{\eta}=0.05\) and \(\lambda=1\), and have a wider negative real range so that additional damped overtones are visible.}
	\label{fig:complex-roots}
\end{figure}

Figure~\ref{fig:complex-roots} visualizes this conclusion by showing the nonzero roots of \(G\) in the complex \(\hat{\omega}\)-plane for \(n=2\).  Panels (a) and (b) show the perfect-fluid subcritical and overcritical cases, respectively: in panel (a) the displayed roots are purely imaginary and represent neutral oscillations, while in panel (b) a real pair appears and the positive real root is the growing instability.  Panels (c) and (d) show the corresponding viscous cases: oscillatory roots move into the left half-plane and become damped modes, while the overcritical panel (d) still contains a positive real root.

\subsection{Near-threshold scaling}

We now describe how the unstable root approaches the marginal point as
\(\beta\) approaches \(\beta_c\) from above.  In the perfect-fluid case,
\(\hat{\eta}=\hat{\zeta}_+=0\), the cubic term in
equation~\eqref{eq:Ghat-smallw} is absent, so the small-root balance is between
the quadratic and quartic terms.  This gives
\eq{
	\hat{\omega}
	=
	\left(
		\frac{4n^3(n+3)^2}{n+1}
	\right)^{1/4}
	(\beta-\beta_c)^{1/2}
	+
	O(\beta-\beta_c),
	\qquad
	\hat{\eta}=\hat{\zeta}_+=0.
}
This is the square-root law obtained by \citet[equation (3.4)]{Miyamoto:2011hr}.

In a viscous fluid with nonzero bulk viscosity, \(\hat{\zeta}>0\), the cubic
coefficient at \(\beta=\beta_c\) is \((n+1)\hat{\zeta}/n\).  The leading
balance is then between the quadratic and cubic terms, and the unstable root
emerges as
\eq{
	\hat{\omega}
	=
	\frac{2}{\hat{\zeta}}
	\left(\frac{n}{n+1}\right)^{3/2}
	(\beta-\beta_c)
	+
	O((\beta-\beta_c)^2),
	\qquad
	\hat{\zeta}>0.
}
If the bulk viscosity vanishes, \(\hat{\zeta}=0\), the cubic coefficient at the
threshold vanishes even when \(\hat{\eta}>0\); the shear-viscous contribution
cancels in this coefficient.  The leading balance is then again
quadratic--quartic, as in the perfect-fluid local scaling.  Thus nonzero bulk
viscosity changes the local onset from square-root to linear, whereas shear
viscosity alone does not change the leading exponent.

\begin{figure}[!htbp]
	\centering
	\includegraphics[width=0.84\linewidth]{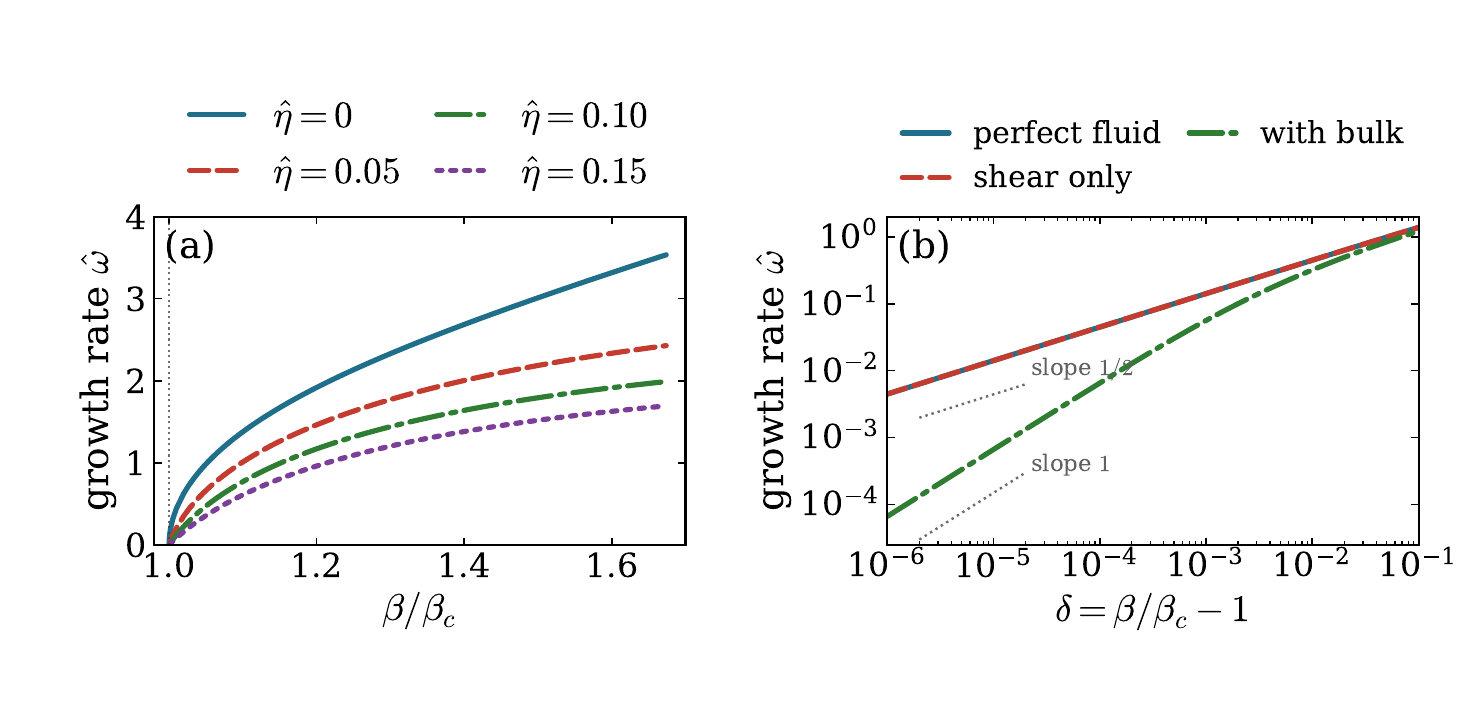}
	\caption{Growth rate \(\hat{\omega}\) for \(n=2\).  Panel (a) shows global curves on linear axes for several \(\hat{\eta}\) (\(\lambda=1\) in viscous cases).  Panel (b) compares the perfect-fluid, shear-only \((\hat{\eta}=0.02,\lambda=0)\), and bulk-viscous \((\hat{\eta}=0.02,\lambda=1)\) curves as functions of \(\delta=\beta/\beta_c-1\) on log--log axes; the dotted segments indicate slopes \(1/2\) and \(1\).}
	\label{fig:marginal-root}
\end{figure}

Figure~\ref{fig:marginal-root} shows the unstable spherical root for \(n=2\).
Panel (a) gives the global curves on linear axes, including the marginal
endpoint \((\beta/\beta_c,\hat{\omega})=(1,0)\).  Panel (b) compares the
perfect-fluid, shear-only, and bulk-viscous results as functions of
\(\delta\coloneqq\beta/\beta_c-1\) on log--log axes; the reference slopes
\(1/2\) and \(1\) show the square-root and linear onsets.

The generally complex subcritical viscous roots are instead displayed in
figure~\ref{fig:complex-roots}.  Appendix~\ref{app:nonspherical-spectrum}
retains the stable side of the perfect-fluid spectrum, whose eigenvalues are
either purely imaginary or real.  There \(|\hat{\omega}|\) is an oscillation
frequency or a growth rate, and only the spherical mode softens to zero.

\FloatBarrier

\section{Thermodynamic interpretation}
\label{sec:thermo}

The instability threshold can also be characterized thermodynamically.  The argument uses only the total energy of a spherical drop at fixed mass and entropy, and therefore contains no viscous coefficient.  It explains why the critical radius is the same in the inviscid and viscous problems, even though their time-dependent eigenvalues differ.  We first derive the general energy criterion and then illustrate it with a polytropic energy landscape.

\subsection{General thermodynamic argument}
\label{sec:thermo-general}

For a spherical drop of radius \(R\), let
\(V(R)=v_{n+1}R^{n+1}\) and \(A(R)=a_nR^n\) denote its volume and surface area,
respectively.  Here \(v_{n+1}\) is the volume of the unit \((n+1)\)-ball and
\(a_n\) is the area of the unit \(n\)-sphere, so that
\(a_n=(n+1)v_{n+1}\) and \(dV/dR=A\).

Let \(U(V,M,S)\) be the internal energy, where \(M\) and \(S\) are the total
mass and entropy.  For the radial perturbations considered here, \(M\) and
\(S\) are fixed, so the first law reduces to \(dU=-p\,dV\).  The total energy
of the drop is
\eq{
	E(R)=U(V(R),M,S)+\sigma A(R),
\label{eq:total-energy}
}

Differentiating equation~\eqref{eq:total-energy} gives
\eq{
	\frac{dE}{dR}
	=
	\left(-p+\frac{n\sigma}{R}\right)A.
\label{eq:energy-first-derivative}
}
At the equilibrium radius \(R={\color{black}R_0}\), \(dE/dR=0\) gives
\(p_0=n\sigma/{\color{black}R_0}\), the Young--Laplace balance in
equation~\eqref{eq:bg}.  Evaluating the second derivative at this equilibrium
gives
\eq{
	\left.\frac{d^2E}{dR^2}\right|_{R={\color{black}R_0}}
	=
		\left(
		-\left.\frac{dp}{dR}\right|_{R={\color{black}R_0}}
		-\frac{n\sigma}{{\color{black}R_0}^2}
		\right){\color{black}A(R_0)}.
\label{eq:energy-second-derivative-equilibrium-raw}
}
Since \(\rho=M/V(R)\), we have \(d\rho/dR=-(n+1)\rho/R\), and hence
\(dp/dR=-(n+1)\rho c_s^2/R\).  At equilibrium this gives
\(\left.dp/dR\right|_{R={\color{black}R_0}}=-(n+1)\rho_0 c_s^2/{\color{black}R_0}\).

It follows that
\eq{
	\left.\frac{d^2E}{dR^2}\right|_{R={\color{black}R_0}}
	=
		\left(
		(n+1)\frac{\rho_0 c_s^2}{{\color{black}R_0}}
		-\frac{n\sigma}{{\color{black}R_0}^2}
		\right){\color{black}A(R_0)}.
\label{eq:energy-second-derivative}
}
Thus the equilibrium is locally stable against uniform radial variations
when \(E''({\color{black}R_0})>0\), equivalently
\((n+1)\rho_0 c_s^2{\color{black}R_0}>n\sigma\).  The marginal condition
\(E''({\color{black}R_0})=0\) gives \((n+1)\rho_0 c_s^2{\color{black}R_0}=n\sigma\), and hence the
critical radius in equation~\eqref{eq:rcrit-visc}.  The dynamical onset
therefore coincides with the change of sign of the energy curvature at the
equilibrium radius.

Appendix~\ref{app:sr-thermo} gives the corresponding special-relativistic
argument, with \(\rho c_s^2\) replaced by \((\mathcal E+p)c_s^2\) in the
curvature formula.

\subsection{Polytropic energy landscape}
\label{sec:thermo-polytrope}

To visualize the energy criterion beyond its local Taylor coefficient, we choose
a polytropic equation of state, which gives an elementary closed form for the
full radius-dependent energy \(E(R)\).

For a polytropic equation of state,
\eq{
	p=K\rho^\gamma,
\label{eq:polytrope-rho-p}
}
where \(K\) is the polytropic constant and \(\gamma\) is the polytropic
exponent.  Using \(\rho=M/V(R)\) and integrating \(dU=-p\,dV\) at fixed \(M\)
and \(S\), we obtain the total energy, up to an additive constant, as
\eq{
	E(R)
	=
	a_n\sigma R^n
	+
	\begin{cases}
	\displaystyle
	\frac{KM^\gamma}{(\gamma-1)v_{n+1}^{\gamma-1}}\,
	R^{(n+1)(1-\gamma)}
	& (\gamma\neq1)\\[6pt]
	-(n+1)KM\log(R/R_{\rm ref})
	& (\gamma=1)
	\end{cases}.
\label{eq:polytrope-energy}
}
Here the reference scale \(R_{\rm ref}\) affects only the suppressed additive
constant.  Differentiating gives, for all \(\gamma\),
\eq{
	\frac{dE}{dR}
	=
		-\frac{(n+1)KM^\gamma}{v_{n+1}^{\gamma-1}}\,
	R^{n-(n+1)\gamma}
	+
	n(n+1)v_{n+1}\sigma R^{n-1}.
\label{eq:polytrope-dE}
}
Setting \(dE/dR=0\) gives
\eq{
	{\color{black}R_0}^{\,1-(n+1)\gamma}
	=
	\frac{n v_{n+1}^{\gamma}\sigma}{KM^\gamma}.
\label{eq:polytrope-Rstar}
}
Equation~\eqref{eq:polytrope-Rstar} is equivalent to the Young--Laplace
condition \(p=n\sigma/{\color{black}R_0}\).

Evaluating the second derivative at \(R={\color{black}R_0}\), and using equation~\eqref{eq:polytrope-Rstar}, we find
\eq{
	E''({\color{black}R_0})
	=
	n a_n\sigma {\color{black}R_0}^{n-2}
	\left[(n+1)\gamma-1\right].
\label{eq:polytrope-Epp}
}
Equivalently, this can be written as \(E''({\color{black}R_0})=n\sigma A({\color{black}R_0}){\color{black}R_0}^{-2}[(n+1)\gamma-1]\), in agreement with the general curvature formula~\eqref{eq:energy-second-derivative}.
Thus the sign of \(E''({\color{black}R_0})\) changes at
\eq{
	\gamma=\gamma_c,
	\qquad
	\gamma_c\coloneqq\frac{1}{n+1}.
\label{eq:polytrope-critical-gamma}
}
At \(\gamma=\gamma_c\), equation~\eqref{eq:polytrope-Rstar} becomes a condition
on the coefficients rather than an equation that selects \({\color{black}R_0}\).  The two
terms in \(dE/dR\) then have the same power of \(R\); if the coefficient
condition is satisfied, \(dE/dR=0\) for every \(R\), and the equilibria form a
flat family.

This result also describes the local shape of the polytropic energy.  If
\(\gamma>\gamma_c\), an equilibrium point is a minimum of \(E(R)\); if
\(\gamma<\gamma_c\), it is a maximum.  Thus equation~\eqref{eq:polytrope-Epp}
directly gives the local stability against radial perturbations.

\begin{figure}[!htbp]
	\centering
		\includegraphics[width=0.48\linewidth]{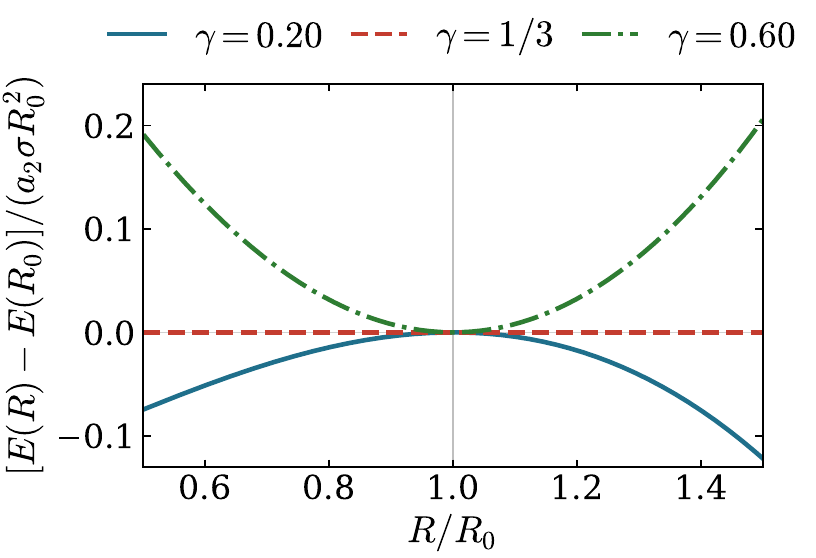}
			\caption{Polytropic energy landscape for \(n=2\), normalized so that the Young--Laplace equilibrium is at \(R/{\color{black}R_0}=1\) for each value of \(\gamma\).  The vertical axis is \([E(R)-E({\color{black}R_0})]/(a_2\sigma {\color{black}R_0}^2)\).  Here \(\gamma_c=1/3\): the curve with \(\gamma<\gamma_c\) has a local maximum, the curve with \(\gamma=\gamma_c\) is locally flat, and the curve with \(\gamma>\gamma_c\) has a local minimum.}
		\label{fig:polytrope}
\end{figure}

Figure~\ref{fig:polytrope} gives the promised global picture for \(n=2\).

\FloatBarrier

Appendix~\ref{app:sr-thermo} shows that the thermodynamically natural
special-relativistic polytrope \(p=Kq^\gamma\), where \(q\) is the conserved
number density, has the same critical exponent \(\gamma_c\).
Appendix~\ref{app:surface-gravity-gamma} compares this surface-tension value
with the corresponding homologous Newtonian-gravity exponent.

\section{Critical radius and diffuse-interface scales}
\label{sec:diffuse}

The sharp-interface treatment used so far assumes a discontinuous density
profile and represents capillarity by a surface tension on an infinitesimally
thin boundary.  It applies when the drop radius \(R\) is much larger than the
finite interface thickness.  We now ask whether the unstable range \(R<{\color{black}R_c}\)
can overlap this range of validity.

The answer is model dependent: in one standard Cahn--Hilliard model the
two ranges do not overlap, while in another they overlap over a parametrically
wide interval.

\subsection{Diffuse-interface preliminaries}

To compare the critical radius with a finite interface thickness, we use the
Cahn--Hilliard diffuse-interface model, in which compressibility and surface
tension arise from the same free-energy functional
\citep{CahnHilliard,Rowlinson,Anderson1998,Jacqmin1999}.

Let \(\rho=\rho(x^I)\) be a density field in \((n+1)\)-dimensional space, and consider the Cahn--Hilliard free energy
\eq{
		F[\rho]
	=
	\int
			\left(
		f(\rho)
		+
			\frac{\varLambda}{2} |\nabla\rho|^2
				\right)\sqrt{g}\,d^{n+1}x,
\label{eq:CH-functional}
}
where \(f(\rho)\) is the homogeneous free-energy density and \(\varLambda>0\) is the gradient coefficient.  For a homogeneous system, \(F_{\rm hom}=Vf(\rho)\) with \(\rho=M/V\), and therefore
\eq{
	p(\rho)=\rho f'(\rho)-f(\rho),
	\qquad
	c_s^2=\rho f''(\rho).
\label{eq:CH-pressure-sound}
}
Thus \(f\) determines the sound speed \(c_s\).

For a planar equilibrium interface with normal coordinate \(y\), the
Euler--Lagrange equation at fixed mass is
\eq{
	f'(\rho)-\varLambda\frac{d^2\rho}{dy^2}=\mu.
}
Here \(\mu\) is the constant chemical potential, or equivalently the
Lagrange multiplier that enforces fixed total mass.  The first integral is
\eq{
		\frac{\varLambda}{2}
	\left(\frac{d\rho}{dy}\right)^2
	=
	W(\rho),
	\qquad
	W(\rho)\coloneqq f(\rho)-\mu\rho+p.
\label{eq:CH-first-integral}
}
Since the surface tension is the excess free energy per unit area and
equation~\eqref{eq:CH-first-integral} makes the homogeneous and gradient
contributions equal, we have
\eq{
	\sigma
	=
	\int_{-\infty}^{\infty}
		\varLambda\left(\frac{d\rho}{dy}\right)^2 dy
		=
		\int_{\rho_{\rm out}}^{\rho_{\rm in}}
		\sqrt{2\varLambda W(\rho)}\,d\rho.
\label{eq:CH-sigma}
}
Here \(\rho_{\rm in}\) and \(\rho_{\rm out}\) denote the two coexisting bulk
densities.
Equations~\eqref{eq:CH-pressure-sound} and \eqref{eq:CH-sigma} show that
\(c_s^2\) and \(\sigma\) are both determined by the diffuse-interface free
energy.  The combination \(\sigma/(\rho c_s^2)\) has dimensions of length, but
its relation to a specific measure of interface thickness is model dependent.
We use the maximum-slope effective thickness
\eq{
	\ell_{\rm eff}
	\coloneqq
	\frac{\rho_{\rm in}-\rho_{\rm out}}{\max |d\rho/dy|}.
\label{eq:ell-eff-def}
}

\subsection{A baseline quartic model}

Let \(s\coloneqq\rho/\rho_0\) denote the normalized density.
As the baseline model, take
\eq{
	f_1(\rho)=a_0\rho^2(\rho_0-\rho)^2,
	\qquad a_0>0.
\label{eq:quartic-free-energy}
}
Figure~\ref{fig:CH-potentials}(a) shows the corresponding normalized
homogeneous free-energy density, with two degenerate minima and a finite
barrier between them.
The two coexisting bulk densities are \(\rho_{\rm out}=0\) and
\(\rho_{\rm in}=\rho_0\).
In this case \(f_1(0)=f_1(\rho_0)=f_1'(0)=f_1'(\rho_0)=0\), so \(\mu=p=0\) and \(W=f_1\).  The sound speed and surface tension are
\eq{
	c_s^2=2a_0\rho_0^3,
	\qquad
	\sigma=\frac{\rho_0^3}{6}\sqrt{2\varLambda a_0},
\label{eq:quartic-cs-sigma}
}
and hence \(\varLambda=36\sigma^2/(c_s^2\rho_0^3)\).

The profile obeys
\eq{
	\frac{d\rho}{dy}
	=
		\sqrt{\frac{2f_1(\rho)}{\varLambda}}
		=
		\sqrt{\frac{2a_0}{\varLambda}}\,\rho(\rho_0-\rho).
\label{eq:quartic-profile-gradient}
}
The maximum gradient occurs at \(\rho=\rho_0/2\).  Using equations~\eqref{eq:ell-eff-def}, \eqref{eq:quartic-cs-sigma} and \eqref{eq:quartic-profile-gradient}, this gives
\eq{
	\ell_{\rm eff}^{(1)}
	=
	\frac{24\sigma}{\rho_0 c_s^2}.
\label{eq:ell-eff-one}
}

Combining equation~\eqref{eq:ell-eff-one} with the critical
radius~\eqref{eq:rcrit-visc} gives
\eq{
	\frac{{\color{black}R_c}}{\ell_{\rm eff}^{(1)}}
	=
	\frac{n}{24(n+1)}
	<
	\frac{1}{24}
	\ll 1.
\label{eq:rc-ell-eff-one}
}
Thus, in the baseline quartic model, the instability does not arise within the
range of validity of the sharp-interface approximation: any drop satisfying
\(R\gg\ell_{\rm eff}^{(1)}\) also has \(R\gg {\color{black}R_c}\) and is therefore stable
against the spherical mode.

\subsection{A shallow-well model}

\begin{figure}[!htbp]
	\centering
	\begin{minipage}[t]{0.39\linewidth}
	\centering
	\includegraphics[width=\linewidth]{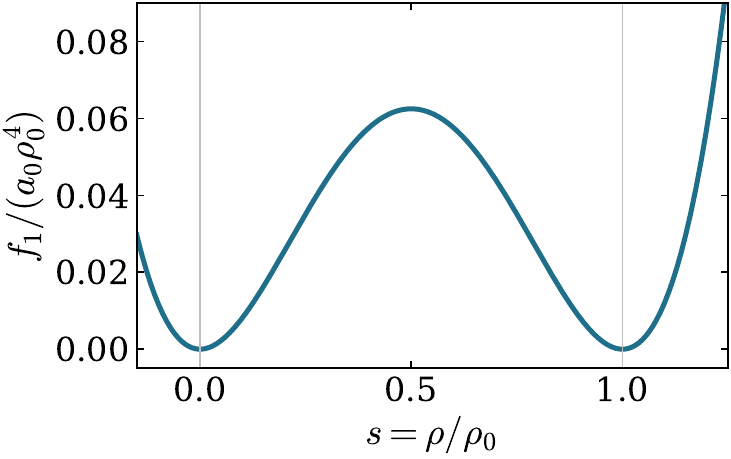}
	\smallskip
	(a) baseline quartic model
	\end{minipage}
	\hfill
	\begin{minipage}[t]{0.39\linewidth}
	\centering
	\includegraphics[width=\linewidth]{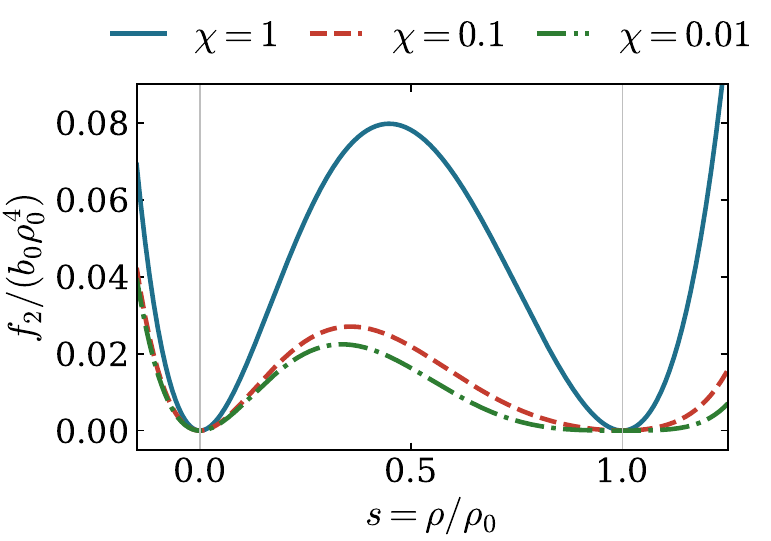}
	\smallskip
	(b) shallow-well model
	\end{minipage}
		\caption{Dimensionless Cahn--Hilliard homogeneous free-energy densities for two distinct models, with \(s=\rho/\rho_0\).  Panel (a) shows the baseline quartic model \(f_1/(a_0\rho_0^4)=s^2(1-s)^2\).  Panel (b) shows the shallow-well model \(f_2/(b_0\rho_0^4)=s^2(1-s)^2[(1-s)^2+\chi]\).  As \(\chi\) decreases, the liquid minimum at \(s=1\) becomes shallow, reducing the sound speed without eliminating the surface-tension integral.}
	\label{fig:CH-potentials}
\end{figure}

To exhibit the modelling freedom explicitly, consider instead the separate
model
\eq{
	f_2(\rho;\chi)
	=
	b_0\rho_0^4 s^2(1-s)^2\left[(1-s)^2+\chi\right],
	\qquad b_0>0,
	\qquad \chi>0.
\label{eq:tunable-free-energy}
}
Figure~\ref{fig:CH-potentials}(b) shows the corresponding normalized
homogeneous free-energy density for several values of \(\chi\).  As \(\chi\)
decreases, the liquid minimum at \(s=1\) becomes progressively shallower,
while the barrier remains finite.
The coexisting bulk densities remain \(\rho_{\rm out}=0\) and
\(\rho_{\rm in}=\rho_0\).  The sound speed and surface tension are
\eq{
	c_s^2=2b_0\rho_0^3\chi,
	\qquad
	\sigma=\rho_0^3\sqrt{2\varLambda b_0}\,\mathcal I(\chi),
	\qquad
	\mathcal I(\chi)\coloneqq\int_0^1s(1-s)\sqrt{(1-s)^2+\chi}\,ds.
\label{eq:tunable-cs-sigma}
}

Equations~\eqref{eq:rcrit-visc} and \eqref{eq:tunable-cs-sigma} give
\eq{
	{\color{black}R_c}
	=
	\frac{n}{n+1}
	\frac{\mathcal I(\chi)}{\rho_0\chi}
	\sqrt{\frac{\varLambda}{2b_0}}.
\label{eq:tunable-rc}
}
To evaluate the maximum-slope thickness defined in
equation~\eqref{eq:ell-eff-def}, consider the equilibrium density profile
across a planar interface, which obeys
\eq{
	\frac{d\rho}{dy}
	=
	\rho_0^2\sqrt{\frac{2b_0}{\varLambda}}\,
	s(1-s)\sqrt{(1-s)^2+\chi}.
\label{eq:tunable-profile-gradient}
}
Finally, take the shallow-well limit \(\chi\to0\).  In this limit,
\(\mathcal I(\chi)\to1/12\), so the surface tension remains finite, while the
maximum of \(s(1-s)\sqrt{(1-s)^2+\chi}\) tends to \(4/27\).
Equations~\eqref{eq:ell-eff-def}, \eqref{eq:tunable-rc} and
\eqref{eq:tunable-profile-gradient} therefore give
\eq{
	\frac{{\color{black}R_c}}{\ell_{\rm eff}^{(2)}}
	\sim
	\frac{n}{81(n+1)}\chi^{-1}
	\longrightarrow\infty.
\label{eq:tunable-ell-separation-asymptotic}
}
Hence there is a parametrically wide range
\(\ell_{\rm eff}^{(2)}\ll R<{\color{black}R_c}\) in which the sharp-interface approximation
is valid and the spherical mode is unstable.  The price is a fine-tuned,
highly compressible liquid phase with \(c_s^2\to0\).

\section{Conclusion}
\label{sec:conclusion}

We have extended the radial instability of a compressible capillary drop from
the perfect-fluid case to a non-relativistic viscous fluid.  The dispersion
relation shows that shear and bulk viscosities change the eigenvalues but not
the onset radius of the radial instability, which is given by
equation~\eqref{eq:rcrit-visc}.  For \({\color{black}R_0}<{\color{black}R_c}\), the corresponding radial
branch contains a positive real root.

Viscosity does change how the growth rate approaches zero at the onset.  The
perfect-fluid root follows a square-root law.  This law remains leading when
only shear viscosity is present, whereas nonzero bulk viscosity changes it to
a linear law.  Thus viscosity changes the dynamics near the threshold without
moving the threshold itself.

This result has a simple thermodynamic interpretation.  For a uniform change
of radius at fixed total mass and entropy, the energy curvature in
equation~\eqref{eq:energy-second-derivative} changes sign at the same radius.
Because viscosity does not enter the equilibrium energy, it can change the
motion without changing this threshold.  For a polytropic equation of state,
the same criterion gives \(\gamma_c=1/(n+1)\), as in
equation~\eqref{eq:polytrope-critical-gamma}.

We finally compared the critical radius with the diffuse-interface thickness
in two Cahn--Hilliard models.  In the baseline quartic model,
equation~\eqref{eq:rc-ell-eff-one} gives
\({\color{black}R_c}/\ell_{\rm eff}^{(1)}<1/24\), so the instability range \(R<{\color{black}R_c}\) does not
overlap the sharp-interface range \(R\gg\ell_{\rm eff}^{(1)}\).  By contrast,
in the separate shallow-well model, the liquid sound speed can be made small
while the surface tension remains finite.  Equation~\eqref{eq:tunable-ell-separation-asymptotic}
then gives \({\color{black}R_c}/\ell_{\rm eff}^{(2)}\to\infty\), so the interval
\(\ell_{\rm eff}^{(2)}\ll R<{\color{black}R_c}\) can be parametrically wide.  This second
result requires a fine-tuned, highly compressible liquid phase.  The relation
between the instability threshold and the interface thickness is therefore
model dependent.

Several directions remain open.  The nonlinear evolution of the radial
instability should be studied for both perfect and viscous fluids in order to
determine the fate of an unstable drop beyond the linear regime.  A full
nonspherical viscous spectrum would show whether other capillary modes become
important near the radial threshold.  Including an exterior fluid would also
allow its pressure, inertia and viscosity to affect the mode spectrum.

The Cahn--Hilliard comparison motivates a dynamical diffuse-interface model in
which the radial stability of a finite-thickness drop can be calculated
directly, rather than inferred from a comparison of length scales.  Another
direction is a relativistic viscous fluid, formulated in a causal theory such
as Israel--Stewart theory \citep{IsraelStewart,HiscockLindblom}.  This would
test whether viscosity changes the relativistic threshold and how it modifies
the dynamics near that threshold.  Together, these problems extend the study
toward nonlinear, finite-thickness and relativistic drop dynamics.

\section*{Acknowledgements}
The author thanks the Research Institute for Mathematical Sciences, an International Joint Usage/Research Center located in Kyoto University, for hospitality and support during part of this work.  This work was supported by JSPS KAKENHI Grant Number 22K03623 and by the Research Institute for Mathematical Sciences.

\appendix

\section*{Appendices}

\section{Mass conservation for a material free boundary}
\label{app:mass-conservation}

Because the instability studied here is spherically symmetric, it is important
to verify that the radial perturbation does not change the total mass of the
drop.  A change of radius might otherwise appear to move the solution to a
background with a different mass rather than describe an admissible
perturbation of a given drop.  The continuity equation is local, whereas the
total mass is an integral over a moving domain, so the motion of the free
boundary must also be taken into account.

We show that the kinematic condition makes the free boundary material and
that, together with the continuity equation and the Reynolds transport
theorem, this implies \(dM/dt=0\).  We give the derivation in covariant form for
arbitrary spatial dimension to match section~\ref{sec:NS} and to justify the
fixed-mass variation used in section~\ref{sec:thermo}.

Let \(D(t)\) denote the region occupied by the drop.  Its free boundary
\(\partial D(t)\) is the surface \(\varphi=0\), with outward unit normal \(n_I\)
defined in section~\ref{sec:NS}.  The total mass is
\eq{
	M(t)
	=
	\int_{D(t)}\rho\sqrt{g}\,d^{d-1}x,
	\qquad
	g\coloneqq\det(g_{IJ}).
\label{eq:app-total-mass}
}

The kinematic condition~\eqref{eq:kinetic-1} states that the boundary is
material, so its outward normal velocity is \(v^I n_I\).  Applying the Reynolds
transport theorem \citep[see e.g.][]{Aris1962} to
equation~\eqref{eq:app-total-mass}, and then using the continuity
equation~\eqref{eq:eoc-1} and the covariant divergence theorem, gives
\eq{
	\frac{dM}{dt}
	&=
	\int_{D(t)}\partial_t\rho\sqrt{g}\,d^{d-1}x
	+
	\int_{\partial D(t)}\rho v^I n_I\,d\varSigma_g
	\nn\\
	&=
	-\int_{D(t)}\nabla_I(\rho v^I)\sqrt{g}\,d^{d-1}x
	+
	\int_{\partial D(t)}\rho v^I n_I\,d\varSigma_g
	\nn\\
	&=
	-\int_{\partial D(t)}\rho v^I n_I\,d\varSigma_g
	+
	\int_{\partial D(t)}\rho v^I n_I\,d\varSigma_g
	=
	0,
\label{eq:app-mass-conservation}
}
where \(d\varSigma_g\) is the induced area element on the free boundary.  Thus mass
conservation follows from the local continuity equation and the material
motion of the boundary.  In particular, the radial perturbation evolves at
fixed total mass, as assumed in section~\ref{sec:thermo}.

\section{Perfect-fluid nonspherical spectrum}
\label{app:nonspherical-spectrum}

\begin{figure}[!htbp]
	\centering
	\includegraphics[width=0.50\linewidth]{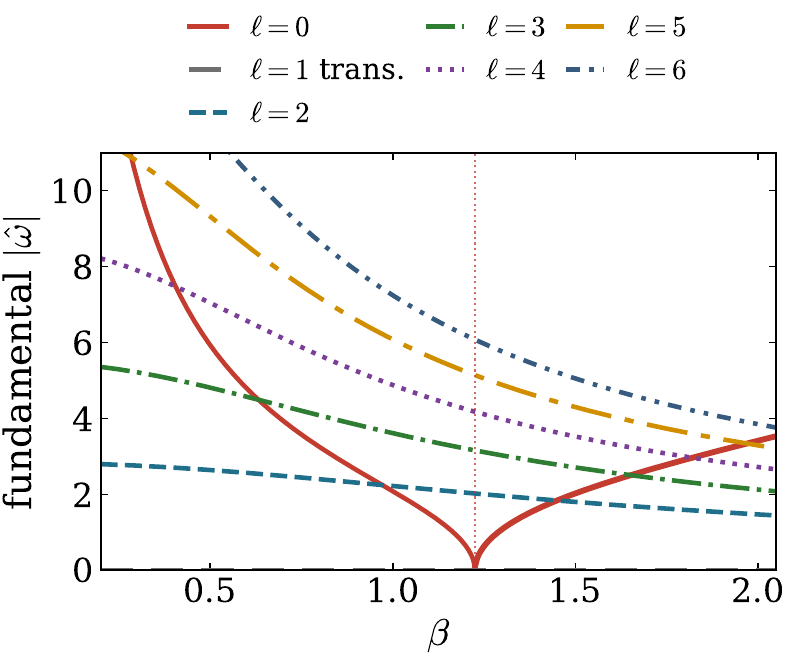}
	\caption{Perfect-fluid fundamental curves for a non-relativistic compressible drop with \(n=2\), computed from the dispersion relation of \citet[Appendix B, equation (B13)]{Miyamoto:2011hr}.  The vertical axis is the fundamental \(|\hat{\omega}|\): for stable modes this is the oscillation frequency, while for the overcritical spherical mode it is the positive real growth rate.  Only the spherical mode \(\ell=0\) reaches zero and turns into an unstable mode; \(\ell=1\) is the translational zero mode and \(\ell\geq2\) are nonspherical capillary oscillations.}
	\label{fig:perfect-nonspherical-spectrum}
\end{figure}

It is useful to see how the nonspherical modes behave as the compressibility is increased.  Figure~\ref{fig:perfect-nonspherical-spectrum} shows the fundamental curves obtained from the perfect-fluid dispersion relation of \citet[Appendix B, equation (B13)]{Miyamoto:2011hr} for a three-dimensional drop, \(n=2\).  Here fundamental means the lowest-\(|\hat{\omega}|\) root in each spherical-harmonic sector \(\ell\).  {\color{black}The plotted interval spans both sides of the spherical threshold.}  The spherical \(\ell=0\) mode softens to zero at the threshold and becomes unstable; \(\ell=1\) is the translational zero mode; and the nonspherical modes \(\ell\geq2\) decrease in frequency but remain oscillatory in the displayed range.  The corresponding viscous nonspherical spectrum will be reported elsewhere.

\section{Special-relativistic thermodynamic threshold}
\label{app:sr-thermo}

This appendix gives the special-relativistic counterpart of the thermodynamic
argument in section~\ref{sec:thermo}, using standard relativistic-fluid
thermodynamics \citep[see e.g.][]{Landau:1987gn} in the capillary-drop setting
of \citet{Miyamoto:2011hr}.

\subsection{General thermodynamic argument}
\label{app:sr-general}

We use units in which the speed of light \(c=1\).  The interior is a
homogeneous relativistic perfect fluid with energy density \(\mathcal E\),
pressure \(p\), total entropy \(S\), and conserved particle number \(N\).  Let
the bulk energy be
\eq{
	U(V,N,S)=\mathcal E V.
\label{eq:sr-bulk-energy-jfm}
}
The radial perturbations considered here keep \(N\) and \(S\) fixed, so
\(dU=-p\,dV\).  Hence the total energy of a spherical drop is
\eq{
	E_{\rm SR}(R)=U(V(R),N,S)+\sigma A(R).
\label{eq:sr-total-energy-jfm}
}
The first derivative is again
\eq{
	\frac{dE_{\rm SR}}{dR}
	=
	\left(-p+\frac{n\sigma}{R}\right)A,
\label{eq:sr-energy-first-jfm}
}
so the equilibrium condition is the Young--Laplace relation
\eq{
	p_0=\frac{n\sigma}{R_0}.
\label{eq:sr-young-laplace-jfm}
}

The second derivative differs from the non-relativistic calculation only
through the compression law.  Since \(d(\mathcal E V)=-p\,dV\),
\eq{
	V\,d\mathcal E=-(\mathcal E+p)\,dV,
	\qquad
	\frac{d\mathcal E}{dR}
	=
	-(n+1)\frac{\mathcal E+p}{R}.
\label{eq:sr-epsilon-radius-jfm}
}
Define the adiabatic relativistic sound speed by
\eq{
	c_s^2=\left(\frac{\partial p}{\partial\mathcal E}\right)_{S/N},
\label{eq:sr-sound-speed-jfm}
}
where the subscript indicates that the specific entropy \(S/N\) is held fixed.
Then
\eq{
	\left.\frac{d^2E_{\rm SR}}{dR^2}\right|_{R_0}
	=
	\frac{{\color{black}A(R_0)}}{R_0}
	\left[
		(n+1)(\mathcal E_0+p_0)c_s^2-p_0
	\right].
\label{eq:sr-energy-curvature-jfm}
}
Thus the non-relativistic factor \(\rho c_s^2\) in
equation~\eqref{eq:energy-second-derivative} is replaced by the relativistic
enthalpy factor \((\mathcal E+p)c_s^2\).  Using
equation~\eqref{eq:sr-young-laplace-jfm}, the marginal radius may be written as
\eq{
	R_{c,{\rm SR}}
	=
	\frac{n\sigma}{(n+1)\mathcal E_0c_s^2}
	\left[1-(n+1)c_s^2\right],
\label{eq:sr-critical-radius-jfm}
}
when \(c_s^2<1/(n+1)\).  If \(c_s^2\geq1/(n+1)\), this
thermodynamic channel has no positive critical radius.
Equation~\eqref{eq:sr-critical-radius-jfm} is identical to the critical radius
obtained from the dynamical perturbation analysis of
\citet{Miyamoto:2011hr}.

\subsection{Polytropic equation of state and energy landscape}
\label{app:sr-polytrope}

For a thermodynamically clean relativistic polytrope, the power law
should be written in terms of the conserved number density
\(q=N/V\):
\eq{
	p=Kq^\gamma
	\qquad
	(K>0).
\label{eq:sr-polytrope-q-jfm}
}
At fixed specific entropy,
\eq{
	d\mathcal E=\frac{\mathcal E+p}{q}\,dq,
\label{eq:sr-epsilon-q-first-law-jfm}
}
which follows from \(dU=-p\,dV\) and \(q=N/V\).  The sound speed is therefore
\eq{
	c_s^2
	=
	\frac{dp/dq}{d\mathcal E/dq}
	=
	\frac{\gamma p}{\mathcal E+p}.
\label{eq:sr-polytrope-sound-jfm}
}
At a marginal equilibrium \(R={\color{black}R_0}\), combining the critical-radius formula
\eqref{eq:sr-critical-radius-jfm} with the Young--Laplace relation
\eqref{eq:sr-young-laplace-jfm} gives
\((n+1)({\color{black}\mathcal E_0}+{\color{black}p_0})c_s^2={\color{black}p_0}\).  Substituting
equation~\eqref{eq:sr-polytrope-sound-jfm} then yields
\([(n+1)\gamma-1]{\color{black}p_0}=0\).  Since \({\color{black}p_0}>0\), the critical exponent is
\(\gamma=\gamma_c=1/(n+1)\), identical to the non-relativistic threshold in
equation~\eqref{eq:polytrope-critical-gamma}.

\begin{figure}[!htbp]
		\centering
		\includegraphics[width=0.48\linewidth]{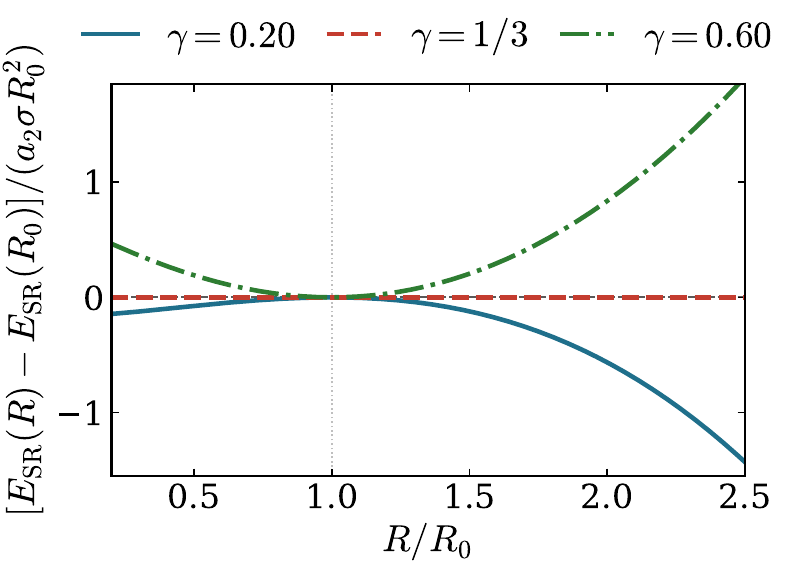}
		\caption{Special-relativistic polytropic energy landscape for
		\(n=2\).  The maximum--flat--minimum transition occurs at
		\(\gamma_c=1/3\).}
	\label{fig:sr-polytrope-landscape}
\end{figure}

To display the full energy landscape, we use \(q=N/V(R)\) and integrate
\(dU=-p\,dV\) at fixed \(S\) and \(N\).  Up to an additive constant, this gives
\eq{
	E_{\rm SR}(R)
	=
	a_n\sigma R^n
	+
	\begin{cases}
	\displaystyle
	\frac{KN^\gamma}{(\gamma-1)v_{n+1}^{\gamma-1}}\,
	R^{(n+1)(1-\gamma)}
	& (\gamma\neq1)\\[6pt]
	-(n+1)KN\log(R/R_{\rm ref})
	& (\gamma=1)
	\end{cases}.
\label{eq:sr-polytrope-energy-landscape-jfm}
}
Here the reference scale \(R_{\rm ref}\) affects only the suppressed additive
constant.  Equation~\eqref{eq:sr-polytrope-energy-landscape-jfm} has the same
radius dependence as its non-relativistic counterpart in
equation~\eqref{eq:polytrope-energy}, with \(M\) replaced by \(N\).  The
argument of section~\ref{sec:thermo-polytrope} therefore applies unchanged: a
Young--Laplace equilibrium is a local maximum for \(\gamma<\gamma_c\), flat at
\(\gamma=\gamma_c\), and a local minimum for \(\gamma>\gamma_c\).
Figure~\ref{fig:sr-polytrope-landscape} illustrates this sign change for
\(n=2\).

\FloatBarrier

\section{Surface tension versus Newtonian gravity}
\label{app:surface-gravity-gamma}

Surface tension and gravity both favor contraction, suggesting a common
stability criterion.  The instability of self-gravitating polytropic stars and
its thermodynamic interpretation are well established
\citep[see e.g.][]{Chandrasekhar1939,Shapiro}.  Here we extend the Newtonian
argument to \((n+1)\) spatial dimensions and compare its critical exponent with
the surface-tension value \(\gamma_c\) in equation~\eqref{eq:polytrope-critical-gamma}.

For a Newtonian self-gravitating homologous model, the radius-dependent energy
may be written as
\eq{
	E_{\rm grav}(R)
	=
	C_{\rm int}R^{-(n+1)(\gamma-1)}
	+
	\begin{cases}
		-\dfrac{C_{\rm grav}}{n-1}R^{-(n-1)}
		& (n>1)\\[1mm]
		C_{\rm grav}\log(R/R_{\rm ref})
		& (n=1)
	\end{cases}.
\label{eq:app-gravity-energy}
}
Here \(C_{\rm int}>0\) sets the internal-energy scale, while
\(C_{\rm grav}>0\) sets the strength of the attractive gravitational energy.
The factor \(1/(n-1)\) makes the \(n\to1\) limit of the first line equal to
the logarithmic second line up to an additive constant.

\begin{figure}[!htbp]
	\centering
		\includegraphics[width=0.52\linewidth]{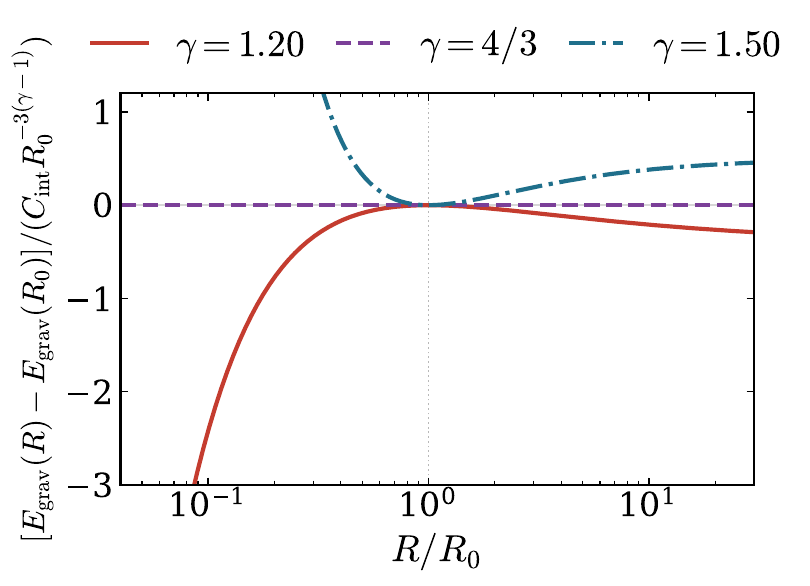}
		\caption{Homologous Newtonian self-gravitating energy landscapes for
		\(n=2\).  The curve with \(\gamma<\gamma_{c,\rm grav}\) has a local
		maximum, the curve with \(\gamma=\gamma_{c,\rm grav}=4/3\) is flat, and
		the curve with \(\gamma>\gamma_{c,\rm grav}\) has a local minimum.}
	\label{fig:gamma-c-surface-gravity}
\end{figure}

The equilibrium condition, valid for \(n\geq1\), is
\eq{
	(n+1)(\gamma-1)C_{\rm int}{\color{black}R_0}^{-(n+1)(\gamma-1)-1}
	=
	C_{\rm grav}{\color{black}R_0}^{-n},
}
and hence
\eq{
	E_{\rm grav}''({\color{black}R_0})
	=
	(n+1)(\gamma-1)C_{\rm int}{\color{black}R_0}^{-(n+1)(\gamma-1)-2}
	\left[(n+1)(\gamma-1)-(n-1)\right].
\label{eq:app-gravity-Epp}
}
The critical exponent is therefore
\eq{
	\gamma_{c,\rm grav}
	\coloneqq
	\frac{2n}{n+1}.
\label{eq:app-gammac-gravity}
}
{\color{black}For \(n=1\), \(\gamma_{c,{\rm grav}}=1\) is a boundary
limit rather than an attained equilibrium: the equilibrium condition above
requires \(\gamma>1\), so no finite equilibrium exists at \(\gamma=1\)
within this homologous model.}
For \(n=2\), equation~\eqref{eq:app-gammac-gravity} gives the familiar
\(\gamma_{c,\rm grav}=4/3\), whereas surface tension gives \(\gamma_c=1/3\).
Figure~\ref{fig:gamma-c-surface-gravity} shows the corresponding
self-gravitating energy landscape for \(n=2\).

\FloatBarrier

\bibliographystyle{plainnat-surname}
\bibliography{refs}

\end{document}